\documentclass{egpubl}
\usepackage{eurovis2026}

\EuroVisShort  

\usepackage[T1]{fontenc}
\usepackage{newtxtext,newtxmath}
\usepackage{amsmath} 
\usepackage[hyphens,spaces,obeyspaces]{url}

\usepackage{cite}  
\BibtexOrBiblatex
\electronicVersion
\PrintedOrElectronic
\ifpdf \usepackage[pdftex]{graphicx} \pdfcompresslevel=9
\else \usepackage[dvips]{graphicx} \fi

\usepackage{egweblnk}

\newcommand{\pheading}[1]{\vspace{4px}\noindent\textbf{#1}}
\usepackage{xcolor}

\newif\ifshowrr
\showrrfalse 
\definecolor{rrcolor}{RGB}{153, 0, 153} 

\newcommand{\rr}[1]{%
  \ifshowrr
    \textcolor{rrcolor}{#1}%
  \else
    #1%
  \fi
}

\title[EG Exploring Proximity Semantics in EDA]%
      {From `Here' to `There': Exploring Proximity Semantics in Multimodal Data Exploration }

\author[Bromley et al.]
{
\parbox{\textwidth}{\centering
Dennis Bromley$^{1}$\orcid{0009-0007-0303-8062}, 
Diana Wang$^{1}$\orcid{0009-0004-7822-0297}, 
and Vidya Setlur$^{1}$\orcid{0000-0003-3722-406X} \\
Tableau Research
}
}

\begin{document}


\maketitle
\begin{abstract}
Modern data exploration tools often struggle to capture the subtleties of analytical intent, especially when users seek patterns that are difficult to specify using traditional query methods or natural language alone. We introduce a multimodal research probe for querying time-series and geospatial data that integrates free-form sketching, natural language, and visual annotations within a unified interaction space. Users articulate queries by sketching trends or spatial paths and augmenting them with annotations and analytical directives grounded in shared spatial and temporal context. The system employs a hybrid architecture combining geometric sketch matching and visual language models (VLMs) to support queries that interleave pattern matching and semantic constraints. Through a preliminary study with 20 participants, we observed recurring interaction patterns in which participants used spatial, temporal, and visual proximity to relate sketches, annotations, and language. Rather than treating these as isolated inputs, participants relied on their relative placement to disambiguate meaning. We analyze these behaviors as evidence for proximity semantics (PS), a form of deictic disambiguation in which meaning is shaped by the closeness of multimodal elements within a shared interaction space. We present PS as a conceptual lens grounded in observed user behavior, and discuss its implications for the design of future multimodal data exploration systems.

\ccsdesc[500]{Human-centered computing~Interaction paradigms}
\ccsdesc[500]{Human-centered computing~Interaction techniques}
\ccsdesc[500]{Human-centered computing~Visualization systems and tools}

\printccsdesc   
\end{abstract}  
\section{Introduction}
Traditional search and query systems depend on users' ability to articulate analytical intent that aligns with the data schema and query language~\cite{Baeza-Yates:1999,manning2008introduction}. This poses challenges when users reason about patterns that are visually intuitive but difficult to express precisely using language-only or query formulations~\cite{li:2016}. Advances in natural language interfaces (NLIs) enable conversational interaction with data~\cite{eviza,datatone,analyza}, but remain largely dependent on verbal formulations and offer limited support for visually grounded reasoning. Sketch-based querying shows that freehand drawing can support the expression of visual patterns~\cite{Sezgin:2001,qetch}; however, these systems typically focus on shape matching and offer limited support for higher-level analytical constraints or multimodal integration. Motivated by the complementary strengths of sketch-based interaction and natural language, our work explores how users combine these modalities within a shared query articulation space. We implemented a research probe that allows users to sketch trends or spatial paths, annotate regions of interest, and write free-form analytical directives on the same canvas (see Figure~\ref{fig:design_probe_interface}). 
\rr{The supported annotation vocabulary (e.g., circles, cross-outs, arrows, and labels) was iteratively developed informed by prior sketch-based systems and common diagrammatic practices, and kept lightweight to allow reinterpretation.} The system supports sketch input to enable a mathematically precise, user-controllable geometric search, while a visual language model (VLM) interprets handwritten text, symbols, and diagrammatic annotations to generate structured database queries. \rr{In a preliminary user study (N=20), we observed recurring interaction patterns in which participants used spatial arrangement and visual proximity to relate sketches, annotations, and text. For example, they placed labels near regions of interest, used arrows to connect text and data, and marked inclusion and exclusion zones directly on the visualization. Rather than presenting a complete interaction paradigm, we treat these observations as empirical grounding for a conceptual contribution. These patterns led us to define \textit{proximity semantics} (PS): a form of deictic disambiguation in which meaning is shaped by the spatial, temporal, and visual closeness of multimodal elements within a shared interaction space. We adopt PS as an analytic lens to characterize these behaviors and inform the design of integrated multimodal analytical interfaces.}

\vspace{-3.8mm}
\section{Related Work}
Our work spans three themes: (1) text-based and semantic search systems, (2) visual query systems, and (3) sketch-based interaction.

\subsection{Text-Based and Semantic Search Systems}
Search systems have diversified to cater to a broad spectrum of search intents~\cite{cimiano:2008,Damljanovic2010NaturalLI,fernandez:2008}. Recent advances in LLMs have expanded the expressive power of search and question answering by enabling conversational querying and programmatic query generation~\cite{openai2023gpt4,anthropic_37_sonnet,lewis-etal-2020-bart}. Despite these advances, language-centric approaches remain constrained by verbal formulations and struggle to express analytical intent that is inherently spatial, visual, or relational. Prior work mapping quantitative properties to linguistic descriptors (e.g., ``spike,'' ``bump'')~\cite{bromley2023difference,bendeck:2024} still relies on predefined vocabularies and offers limited support for visually grounded reference or deictic reasoning. \rr{While prior systems support referring expressions (e.g., ``this region''), they typically resolve them through explicit identifiers or structured context rather than \textit{spatial or temporal proximity}. As a result, they have limited ability to capture the implicit, spatially grounded references that arise in sketch- and annotation-based interaction.}
 
\subsection{Visual Query Systems}
Visual query systems leverage direct manipulation, graphical metaphors, and visual encodings to lower barriers to data exploration~\cite{Batini1991VisualQS,siddiqui2016zenvisage}. Systems such as Visage~\cite{Derthick:1997} and time-box queries~\cite{Hochheiser2004DynamicQT} enable users to specify constraints through spatial selection and interaction, supporting iterative and exploratory workflows. \rr{These systems support deictic reference through brushing or bounding boxes, but such references are typically tied to explicit selection primitives rather than flexible, user-defined annotations.} Hybrid approaches have integrated graphical interaction with language or attribute selection~\cite{keim:1992}, but typically treat modalities as parallel input channels rather than as elements within a shared semantic space. As a result, many visual query systems struggle with open-ended analytical expression involving ambiguous references, relational constraints, or qualitative notions such as ``near'' or ``not this region.'' \rr{In particular, they offer limited capability for interpreting references that are implied through the relative placement of marks, text, and data rather than through explicit selection or linking mechanisms.}

\subsection{Sketch-based Interaction}
Sketch-based interaction has long been studied as a means of facilitating informal, expressive input where rigid syntax is limiting~\cite{Landay1995InteractiveSF,Sezgin:2001,hammond:2007}. In data-centric contexts, sketching has been used for shape-based querying and pattern matching in time-series and spatial data~\cite{qetch,siddiqui2021shapesearch,sensor}. More recent systems extend sketching toward exploratory analysis, sensemaking, and communication, including narrative visualization and hypothesis-driven exploration~\cite{sketchstory:2013,activeink:2019,Lin2023InkSightLS}. 
\rr{Recent work by Yen et al.~\cite{yen:2025} further explores how free-form sketches, annotations, and spatial relationships can be interpreted by AI models to iteratively guide system behavior in a shared workspace. Although the system incorporates multimodal inputs, the focus is on mapping sketch-based annotations to code edits within a localized programming context, rather than supporting broader multimodal queries that combine sketch, language, and deictic reference within a shared interaction space.} Emerging work such as SketchGPT~\cite{sketchgpt} highlights the potential of combining sketch and language, but the role of spatial proximity and shared context in grounding analytical meaning remains underexplored.

\rr{More broadly, prior work on deictic interaction (e.g., pointing, labeling, and gestural reference) has shown that users rely on spatial and temporal context to disambiguate meaning, but these mechanisms are typically treated as explicit signals (e.g., arrows, selections, gestures) rather than as emergent relationships between colocated multimodal elements.} Our work builds on these efforts by examining how users combine sketching, annotation, and language when they coexist in a shared interaction space. Through a multimodal design probe and user study, we observed interaction patterns in which users employed spatial, temporal, and visual proximity to relate sketches, annotations, and language. Reflecting on these patterns led us to formalize PS as a lens for reasoning about multimodal analytical query systems.

\section{Design Probe for Exploring Multimodal Query Articulation}
\begin{figure}[ht]
\centering
\includegraphics[width=0.83\linewidth]{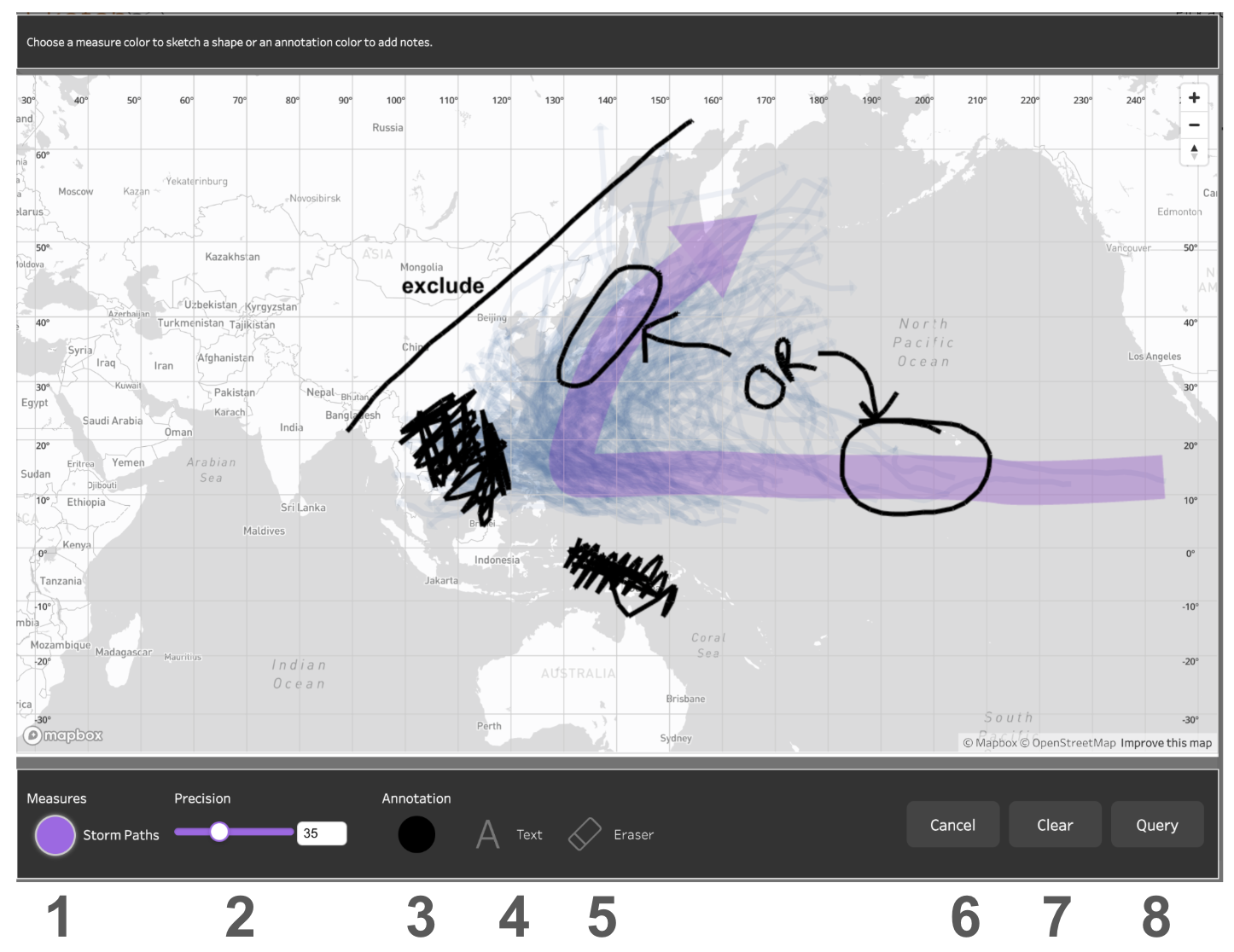}
\caption{Design probe interface showing trend sketch (purple) and annotations (black). 1. Trend sketch; 2. Precision control; 3. Annotation tool; 4. Text input; 5. Eraser; 6. Cancel; 7. Clear; 8. Query.}
\label{fig:design_probe_interface}
\end{figure}

Our design probe interface is shown in Figure~\ref{fig:design_probe_interface}. The probe explores how users combine geometric sketch-based search~\cite{qetch} with VLM-based interpretation of free-form annotations, with the goal of understanding emergent interaction patterns. The two interpretation pipelines operate separately but can be used in parallel, with their results intersected for final output. Brief implementation details are provided here with additional details in the supplemental materials.

\subsection{Geometric Trend Search}

\begin{figure}[ht]
  \centering
  \includegraphics[width=0.84\linewidth]{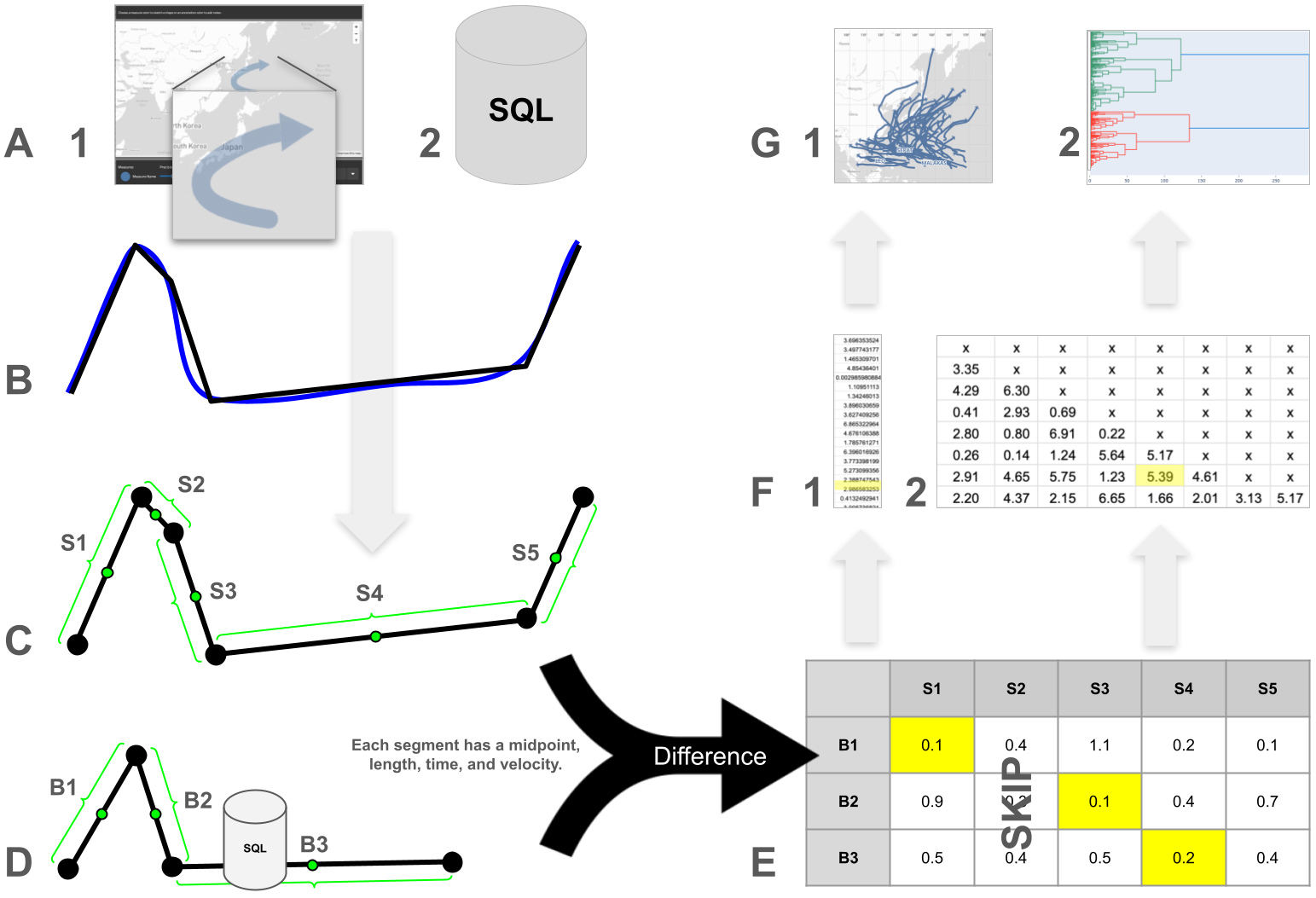}
  \caption{Overview of the probe's geometric processing pipeline.}
  \label{fig:geom_sketch_arch}
\end{figure}

The geometric trend search interprets a user’s sketch and compares it to trends in the database. User-controllable \rr{comparison penalties} (e.g., precision, time invariance, sub-shape matching) \rr{allow fine-grained control over how sketches are matched (e.g., setting the time penalty to zero enables matches regardless of temporal alignment)}. Figure~\ref{fig:geom_sketch_arch} illustrates this process. (A) A signal originates from either a user sketch (A.1) or the database (A.2). (B) Signals are simplified using the Douglas-Peucker algorithm~\cite{douglas1973algorithms}. (C) Segment properties (e.g., length, angle, position) are extracted. (D) Database signals are \rr{similarly} processed. (E) Signals are compared on a per-segment, \rr{per-property basis, with differences weighted by user-defined penalties to compute a final similarity score}. If the input is a single sketch, the comparison yields a $\text{1} \times \text{N}$ table (F1), which is sorted to return best matches (G1). If applied across the dataset, it produces an $\text{N} \times \text{N}$ matrix (F2) used for analyses such as clustering (G2).

\subsection{VLM-Interpreted Free-Form Annotations}

Users can annotate data using natural language (e.g., ``only storms after 1970''), symbols (e.g., arrows, \textsc{OR}), and glyphs such as cross-outs, circles, and boundary lines. While we provide default interpretations, users may repurpose these marks. The annotated image is sent to a VLM (Claude 3.7 Sonnet~\cite{anthropic_37_sonnet}) along with prompts for task framing, schema grounding, visual context, glyph interpretation, and SQL generation. The VLM produces \rr{an annotation-derived SQL query}, whose results are intersected with those of the geometric search. \rr{We separate the VLM and geometric pipelines because they support different forms of control: annotations require flexible, high-level interpretation suited to language models, while trend matching relies on precise, user-tunable numeric parameters. This separation allows each modality to leverage its strengths while enabling unified query execution through result intersection.}

\section{Preliminary Evaluation and Reflection}
To evaluate our design probe, we conducted 45-minute interview sessions with 20 participants, including data scientists, sales consultants, and product managers. 
\rr{Participants were recruited through internal mailing lists and had prior experience with data visualization, though not necessarily with sketch-based querying.}

\begin{itemize}
\item \textit{Introduction} [$\sim$5 min.]: Participants were introduced to the probe. 
\rr{While no formal training was provided, we demonstrated the available tools (e.g., sketching, annotation, and text input) to ensure awareness without prescribing usage strategies or exposing the probe datasets.}

\item \textit{Task Completion} [$\sim$25 min.]: Participants worked with two datasets: \href{https://www.kaggle.com/datasets/kaggle/us-baby-names}{baby names} and \href{https://www.kaggle.com/datasets/chriszhengao/cma-best-track-data}{storm paths}, completing five predefined tasks.

\item \textit{Line Chart -- Baby Names} [$\sim$10 min.]: Participants searched for names matching nuanced temporal patterns (e.g., peaks or resurgence).

\item \textit{Map -- Storm Paths} [$\sim$10 min.]: Participants identified geographic patterns in storm trajectories using shape-based input and region filtering.

\item \textit{Open-Ended Task} [$\sim$5 min.]: Participants explored the system freely, \rr{allowing us to observe emergent behaviors beyond predefined tasks.}

\item \textit{Feedback Session} [$\sim$10 min.]: Semi-structured interviews captured participant impressions. 
\rr{Sessions were recorded and transcribed for analysis.}

\item \textit{System Usability Scale (SUS)}: Participants completed the SUS survey~\cite{brooke1996sus}. \rr{We selected SUS as a lightweight, widely used measure to provide a baseline assessment of perceived usability, complementing our qualitative analysis rather than serving as the primary evaluative instrument.} 
\end{itemize}

\pheading{Analysis method.} 
\rr{We conducted a thematic analysis of transcripts and interaction logs. Two authors independently identified recurring patterns in how participants combined sketches, annotations, and text, then iteratively grouped them into higher-level categories. Disagreements were resolved through discussion. This process revealed a recurring pattern in which participants used spatial, temporal, and visual proximity to relate multimodal elements.}

\begin{figure}[ht]
\centering
\includegraphics[width=1.0\linewidth]{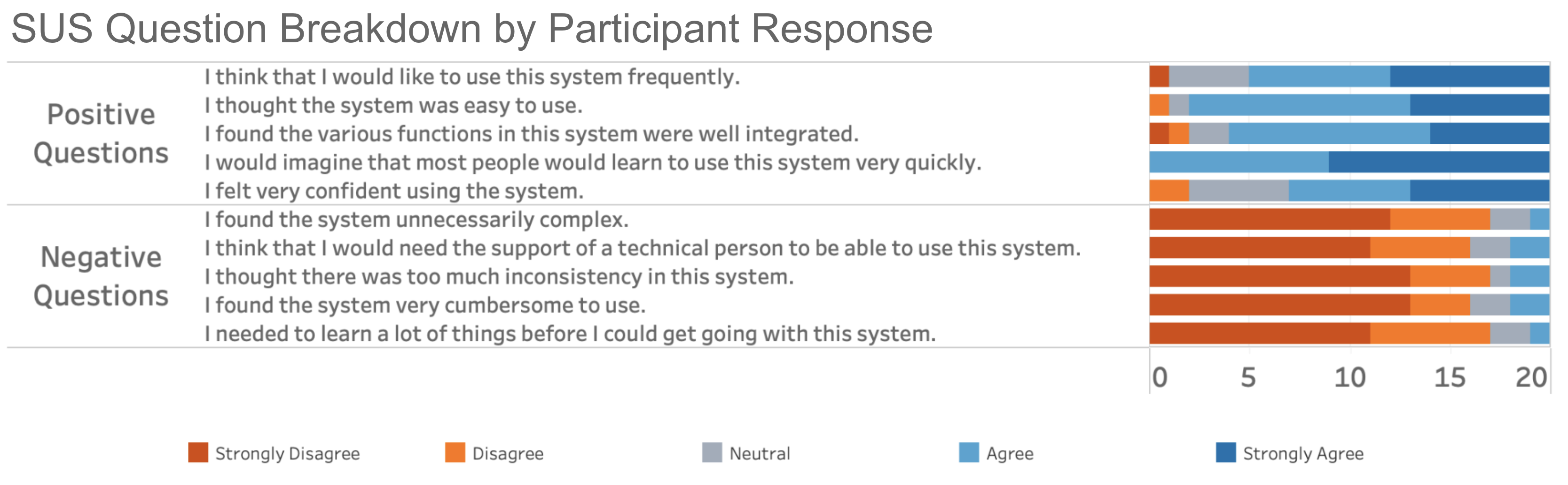}
\caption{Breakdown of participant responses to SUS questions.}
\label{fig:sus_scores}
\end{figure}

Participants found the probe expressive for exploring complex data trends, with an average SUS score of $81$ ($\geq 68$ indicates good usability~\cite{Bangor_SUS_adjective}). They appreciated the intuitive sketching interface and integration of annotations, while suggesting improvements in query interpretation, feedback, and onboarding. \rr{17 of 20 participants used spatial proximity to convey meaning, often placing labels or glyphs adjacent to regions and referring to them with terms such as ``this'' or ``here.'' In addition, participants frequently combined proximity with explicit marks: arrows were commonly used to connect text and data regions, while circles and cross-outs were used to indicate inclusion and exclusion. However, these marks were rarely used in isolation; rather, their interpretation depended on spatial placement relative to candidate referents.} 
\rr{For example, P11 labeled and referenced regions textually, forming multimodal queries through spatial grouping. However, three participants relied primarily on a single modality, producing less precise or more ambiguous queries.}

Through analysis, we observed interaction patterns in which participants relied on spatial, temporal, and visual proximity to relate sketches, annotations, and language. 
\rr{These relationships were often established implicitly through placement rather than explicit connectors. Proximity functioned as a mechanism for disambiguating references, particularly when multiple candidate referents were present.} 
These observations motivated us to examine proximity as a mechanism for multimodal query articulation. In the following section, we introduce \textit{proximity semantics} as a lens for describing this form of deictic disambiguation.

\section{Proximity Semantics}

\begin{figure}[ht]
  \centering
  \includegraphics[width=.83\linewidth]{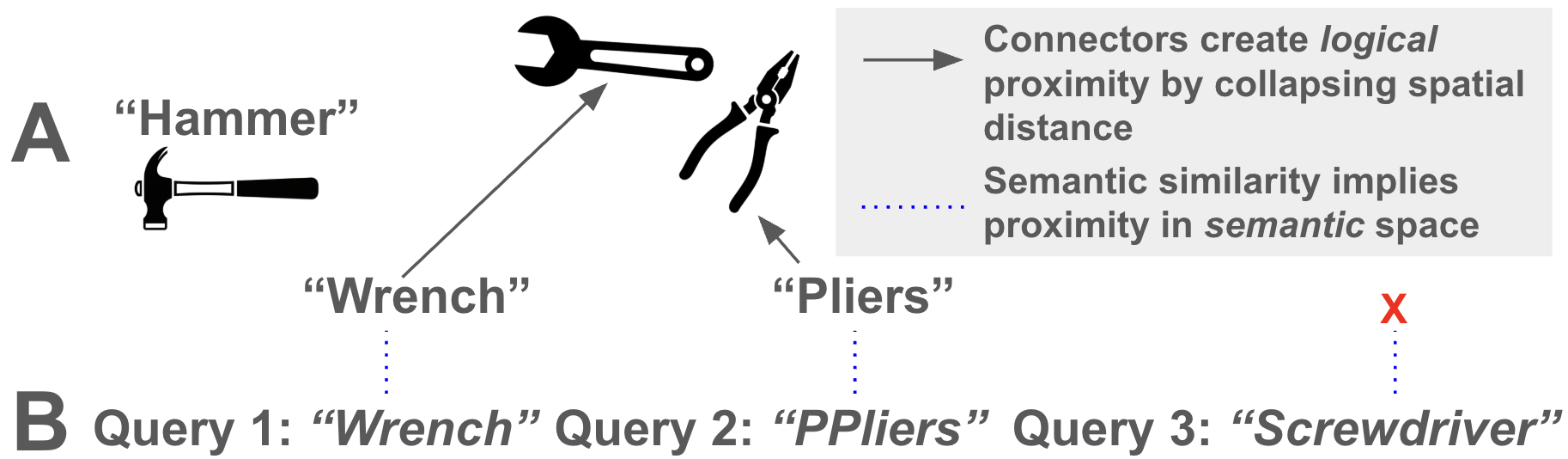}
  \caption{\textbf{A.} PS Examples. Simple spatial proximity (labeled hammer) and logical spatial proximity (wrench and pliers references). \textbf{B.} PS chains linking queries to objects across semantic and canvas spaces; red `x' indicates no semantic match.}
  \label{fig:proximity_semantics_explanation}
\end{figure}

\rr{\textit{Proximity Semantics (PS)} is a reference-resolution framework in which relationships are established through chains of proximity bindings across measurable spaces (e.g. Euclidean, semantic, and temporal), yielding resolution sets of zero, one, or multiple candidates; it does not cover, for example, identification by property (e.g. ``the largest'') or shared knowledge (e.g. ``that thing we saw''). PS thus provides a single formalism for the deictic relationships that underlie common authoring conventions such as labels, arrows, and footnotes.  PS has two primitives: \textit{proximity bindings} and \textit{connectors}. Consider Figure~\ref{fig:proximity_semantics_explanation}.A; a proximity binding is a referential relationship between two elements (e.g. a hammer and its label ``Hammer'') whose measured distance in a common space falls within an interpretive threshold i.e., the two objects appear `close' together. A \textit{connector} such as an arrow can collapse distance within a space; this \textit{logically} places two items next to each other, reducing complex referential relationships to simple proximity relationships, regardless of the length of the connector. For example, the label ``Wrench'' is spatially closest to the hammer icon, but the label's arrow collapses its indicated distance, placing the label \textit{logically} closest to the wrench icon and establishing the proximity binding.}

\rr{A PS binding chain can cross multiple measurable spaces. In Figure~\ref{fig:proximity_semantics_explanation}.B, the exact semantic match (blue dotted line) between \textit{Query 1} and Figure~\ref{fig:proximity_semantics_explanation}.A's wrench label (both ``Wrench'') renders them proximal in \textit{semantic} space. Together with the \textit{logical} spatial proximity of Figure~\ref{fig:proximity_semantics_explanation}.A's wrench and label, the PS proximity chain from \textit{Query 1} to wrench icon is established. \textit{Query 2} follows a similar chain despite the misspelling because the semantic distance between ``PPliers'' and ``Pliers'' falls within our interpretive threshold, establishing semantic proximity and thus the full PS chain. Finally, \textit{Query 3} is not semantically proximal to any of Figure~\ref{fig:proximity_semantics_explanation}.A's labels and the PS chain is not established.
}


\rr{
Given this, we now consider PS-interpretations of common deictic scenarios. For example, the arrow deixis in Figure \ref{fig:proximity_semantics_example1}.A adheres to Kaplan's \cite{kaplan1989demonstratives} formalism of \textit{demonstratives}, \textit{demonstrations}, and \textit{demonstrata}. Within the PS conceptualization however, this becomes a simple proximity chain where the arrow connectors collapse the distance between the words and glyphs, logically placing `this' next to the circle around Japan and `here' next to the exclusion cross-out in the Pacific, establishing disambiguated proximity-based deixis despite word:glyph distances in canvas space.}

\subsection{PS Query Space Definition and Query Tool Requirements}
\rr{
We define a PS Query Space as the set of queries whose correct interpretation depends on proximity-based deictic disambiguation. These are cases in which multiple plausible referents exist and cannot be resolved without establishing a PS chain through one or more measurable spaces. A PS query tool must therefore be capable of using single- or multi-space proximity bindings and connectors to establish a PS chain and resolve semantic ambiguity.
}



\subsection{PS Query Tool Example}
\begin{figure}[ht]
  \centering
  \includegraphics[width=.82\linewidth]{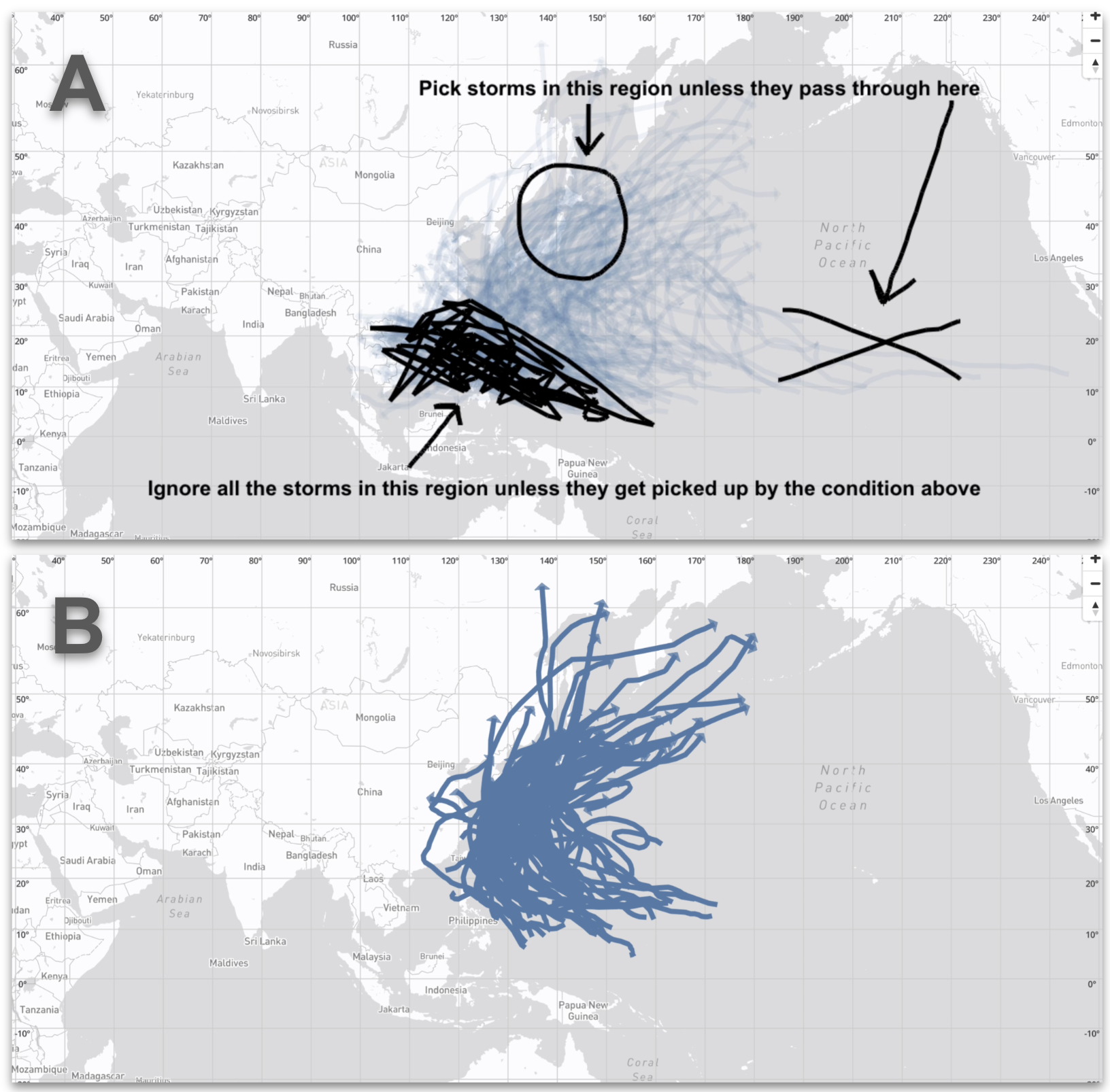}
  \caption{\textbf{A.} A geospatial PS query. Multiple deictic references require proximity-based disambiguation for proper semantic interpretation. \textbf{B.} Results of query interpretation.
}
  \label{fig:proximity_semantics_example1}
\end{figure}

\rr{
We revisit the design probe with these PS definitions and requirements in mind. Figure~\ref{fig:proximity_semantics_example1} shows a multimodal query in which geospatial data is filtered using natural language and annotations. The semantics of the query cannot be determined from any single modality. Instead, meaning emerges by resolving the PS chain that links textual references, glyphs, and data regions within the shared visual space.
}


\section{Conclusion}
Motivated by interaction patterns observed in our design probe, we \rr{introduce } the notion of PS and provide \rr{examples} of both general multimodal interaction and natural language-visual querying contexts. We derive design recommendations for systems that support proximity-based \rr{query} interpretations, demonstrate a research probe capable of executing these queries\rr{, and argue that PS is a unifying framework for deictic reference and a common design vocabulary for multimodal interfaces}. Although PS represents a bounded interaction space, our findings suggest that spatial, temporal, and syntactic forms of proximity are meaningful mechanisms through which users express analytical intent.

\newpage

\bibliographystyle{eg-alpha-doi} 
\bibliography{main}       


\end{document}